\documentstyle[12pt]
{article}
\def \el {{\ell}}
\def \KK {{\cal  K}}
\def \K {{\rm K}}

\newcounter{subequation}[equation]

\makeatletter

\def\thesubequation{\theequation\@alph\c@subequation}
\def\@subeqnnum{{\rm (\thesubequation)}}
\def\slabel#1{\@bsphack\if@filesw {\let\thepage\relax
   \xdef\@gtempa{\write\@auxout{\string
      \newlabel{#1}{{\thesubequation}{\thepage}}}}}\@gtempa
   \if@nobreak \ifvmode\nobreak\fi\fi\fi\@esphack}
\def\subeqnarray{\stepcounter{equation}
\let\@currentlabel=\theequation\global\c@subequation\@ne
\global\@eqnswtrue \global\@eqcnt\z@\tabskip\@centering\let\\=\@subeqncr

$$\halign to \displaywidth\bgroup\@eqnsel\hskip\@centering
  $\displaystyle\tabskip\z@{##}$&\global\@eqcnt\@ne
  \hskip 2\arraycolsep \hfil${##}$\hfil
  &\global\@eqcnt\tw@ \hskip 2\arraycolsep
  $\displaystyle\tabskip\z@{##}$\hfil
   \tabskip\@centering&\llap{##}\tabskip\z@\cr}
\def\endsubeqnarray{\@@subeqncr\egroup
                     $$\global\@ignoretrue}
\def\@subeqncr{{\ifnum0=`}\fi\@ifstar{\global\@eqpen\@M
    \@ysubeqncr}{\global\@eqpen\interdisplaylinepenalty \@ysubeqncr}}
\def\@ysubeqncr{\@ifnextchar [{\@xsubeqncr}{\@xsubeqncr[\z@]}}
\def\@xsubeqncr[#1]{\ifnum0=`{\fi}\@@subeqncr
   \noalign{\penalty\@eqpen\vskip\jot\vskip #1\relax}}
\def\@@subeqncr{\let\@tempa\relax
    \ifcase\@eqcnt \def\@tempa{& & &}\or \def\@tempa{& &}
      \else \def\@tempa{&}\fi
     \@tempa \if@eqnsw\@subeqnnum\refstepcounter{subequation}\fi
     \global\@eqnswtrue\global\@eqcnt\z@\cr}
\let\@ssubeqncr=\@subeqncr
\@namedef{subeqnarray*}{\def\@subeqncr{\nonumber\@ssubeqncr}\subeqnarray}

\@namedef{endsubeqnarray*}{\global\advance\c@equation\m@ne%
                           \nonumber\endsubeqnarray}

\makeatletter \@addtoreset{equation}{section} \makeatother
\renewcommand{\theequation}{\thesection.\arabic{equation}}

\def \ci {\cite}
\newcommand{\rf}[1]{(\ref{#1})}
\def \la {\label}
\def \const {{\rm const}}
\catcode`\@=11

\newcount\hour
\newcount\minute
\newtoks\amorpm \hour=\time\divide\hour by 60\minute
=\time{\multiply\hour by 60 \global\advance\minute by-\hour}
\edef\standardtime{{\ifnum\hour<12 \global\amorpm={am}%
        \else\global\amorpm={pm}\advance\hour by-12 \fi
        \ifnum\hour=0 \hour=12 \fi
        \number\hour:\ifnum\minute<10
        0\fi\number\minute\the\amorpm}}
\edef\militarytime{\number\hour:\ifnum\minute<10 0\fi\number\minute}

\def\draftlabel#1{{\@bsphack\if@filesw {\let\thepage\relax
   \xdef\@gtempa{\write\@auxout{\string
      \newlabel{#1}{{\@currentlabel}{\thepage}}}}}\@gtempa
   \if@nobreak \ifvmode\nobreak\fi\fi\fi\@esphack}
        \gdef\@eqnlabel{#1}}
\def\@eqnlabel{}
\def\@vacuum{}
\def\marginnote#1{}
\def\draftmarginnote#1{\marginpar{\raggedright\scriptsize\tt#1}}
\def\draft{
        \pagestyle{plain}
        \overfullrule=2pt
        \oddsidemargin -.5truein
        \def\@oddhead{\sl \phantom{\today\quad\militarytime} \hfil
        \smash{\Large\sl DRAFT} \hfil \today\quad\militarytime}
        \let\@evenhead\@oddhead
        \let\label=\draftlabel
        \let\marginnote=\draftmarginnote
        \def\ps@empty{\let\@mkboth\@gobbletwo
        \def\@oddfoot{\hfil \smash{\Large\sl DRAFT} \hfil}
        \let\@evenfoot\@oddhead}

\def\@eqnnum{(\theequation)\rlap{\kern\marginparsep\tt\@eqnlabel}%
        \global\let\@eqnlabel\@vacuum}  }

\renewcommand{\rf}[1]{(\ref{#1})}
\renewcommand{\theequation}{\thesection.\arabic{equation}}
\renewcommand{\thefootnote}{\fnsymbol{footnote}}

\def\appendix#1{
  \addtocounter{section}{1}
  \setcounter{equation}{0}
  \renewcommand{\thesection}{\Alph{section}}
  \section*{Appendix \thesection\protect\indent \parbox[t]{11.15cm}
  {#1} }
  \addcontentsline{toc}{section}{Appendix \thesection\ \ \ #1}
  }
\voffset =
-3truecm
\def \tx {\textstyle}
\def \bi{\bibitem}

\def \ov {\over}
\def \ha {\textstyle { 1\ov 2}}

 \def\ep {\epsilon}

\def \del { \partial}

\def \p {\phi}
\def \ee {\epsilon}
\def \te {\tilde \epsilon}

\def\pd{\partial}

\def\m{\mu}

\def\g{\gamma}
\def\G{\Gamma}

\def\r{\rho}

\def\s{\sigma}

\def\te{\theta} 
\def\ta {\tau}

\def\p{\phi}

\def\ep{\epsilon}

\def \foot{ \footnote}
\def\be{\begin{equation}}
\def\ee{\end{equation}}

\def \ci {\cite}
\def \g {\gamma}
\def \G {\Gamma}
\def \k {\kappa}

\def \m {\mu}

\def \W{{\cal W}}
\def \eps {\epsilon}
\def \ha{{
 { 1 \ov 2}} }
\def \de{{
{ 1 \ov 9}} }
\def \si{{
 { 1 \ov 6}} }
\def \fo{{
{ 1 \ov 4}} }
\def \ei{{
{ 1 \ov 8}} }

\def \rr {{\bar \rho}}

\def\te{\theta}
\def\g{\gamma}

\date{}
\begin{document}

\begin{titlepage}

\begin{center}
\hfill MCTP-00-14\\
\hfill  OHSTPY-HEP-T-01-001\\

\vskip 2.5 cm \vskip 1 cm

{\Large \bf  3-Branes on Spaces with $R\times S^2\times S^3$ Topology}

\vskip .7 cm

\vskip 1 cm

{\large L.A. Pando Zayas$^a$ and A.A. Tseytlin$^{b,c,}$\footnote{Also at
Lebedev
Institute, Moscow.}}\\

\end{center}

\vskip 1cm \centerline{\it ${}^a$ Michigan Center for Theoretical
Physics}
\centerline{ \it Randall Laboratory of Physics, The University of
Michigan}
\centerline{\it Ann Arbor, MI 48109-1120, USA} \centerline{\tt
lpandoz@umich.edu}

\vskip 0.4 cm

\centerline{\it ${}^b$ Department of Physics, The Ohio State
University,}
\centerline{\it 174 West 18th Avenue, Columbus, OH 43210-1106, USA}
\centerline{\tt
tseytlin.1@osu.edu }

\vskip 0.4 cm

\centerline{\it  ${}^c$ Theoretical Physics Group, Blackett Laboratory,}
\centerline{\it Imperial College, London SW7 2BZ, U.K.}

\vskip 1.5 cm

\begin{abstract}
We study supergravity solutions representing D3-branes with 
transverse 6-space having $R\times
S^2\times S^3$ topology.
We consider regular and fractional D3-branes on
a natural one-parameter extensions of the standard 
Calabi-Yau metrics on the singular and resolved conifolds.
After imposing a ${\bf Z}_2$ identification on  an angular 
coordinate these  generalized  ``6-d conifolds" are
nonsingular spaces. The backreaction of D3-branes
creates a curvature singularity that coincides with a horizon.
In  the presence of fractional D3-branes the solutions 
are similar to  the original  ones in hep-th/0002159, hep-th/0010088: 
the  metric has a naked repulson-type singularity located behind 
the radius where the 5-form flux vanishes.
The semiclassical behavior of the Wilson loop suggests that the
corresponding  gauge theory duals  are confining.
\end{abstract}

\end{titlepage}

\setcounter{page}{1}
\renewcommand{\thefootnote}{\arabic{footnote}}
\setcounter{footnote}{0}

\def \N{{\cal N}}
\def \ov {\over}

\section{Introduction}

 One  fruitful approach
to generalize the original AdS/CFT correspondence \ci{ads1}
 to  more ``realistic"
gauge theories with less supersymmetries  is based on
 considering D3-branes on conifold
singularities \cite{moo,kehagias,kw1,mr}. To get non-conformal theories
 one  may add  \cite{kg,kn}
``fractional" \ci{gim}
 D3-branes.

Recently, exact supergravity solutions representing such configurations
were
constructed with
  the 6-d space
transverse to the D3-brane  being  a  conifold \cite{KT}, its
deformation
\cite{KS} and its resolution \cite{pt}.

The deformed conifold \ci{KS}  and resolved conifold \ci{pt}
backgrounds  are two
different (deformation $\ep$ and resolution $a$)  one-parameter
generalizations  of
the  conifold \ci{KT}  one. The three solutions coincide for large
values of the
radial coordinate $\r$, or in the UV in the language of
 gauge theory duals. However,  the small-distance or  IR
behavior is different in each case. In particular, the conifold and the
resolved
conifold solutions have naked singularities at finite $\r$, while (a
special  case
of) the deformed conifold solution is  regular \ci{KS}.

 In view of the  interest  of these
solutions both from
 the supergravity and the gauge theory points of view
(see   \ci{polg,gub,cv1,cv2,papa2,buch,MN,lerd,herz,aha} for some recent related
work) it is
important to study  their various generalizations. That may help to
clarify  which
of their features are truly universal, in particular, regarding
singularities and
IR behavior.

Here we shall point out that there exists a  very natural one-parameter
extension
of  the three solutions of \ci{KT,KS,pt}. The key observation is that
the  most
general  Ricci-flat K\"ahler 6-d metrics on the  three  conifolds
\ci{candelas,minasian} (with non-trivial dependence on  radial direction
only)
contain \ci{pt,papa2} one extra parameter (called $b$ below). This
parameter
 was set equal to zero in the previous
discussions  of D3-branes on conifolds.\foot{For $b\not=0$ the  conifold
metric is
no longer that of a cone over $T^{1,1}$.} These metrics have the same $R
\times S^2
\times S^3$ topology as their $b=0$  limits, in particular, 
 they again
 allow introduction of   fractional 3-branes (D5-branes wrapped
over 2-cycle). 
For $b\not=0$  the 
 conifold and resolved conifold  metrics have regular curvature;
it is possible 
to  remove their  bolt-type  singularity by  
imposing a ${\bf Z}_2$ identification on the coordinate
of the $U(1)$ fiber thus  obtaining non-singular 6-d metrics.
In contrast, the $b\not=0$ generalization of the (regular) 
deformed conifold metric \ci{candelas,minasian}
  has  a curvature singularity at the origin.

As was shown in \ci{papa2}, for any  zero or non-zero
 value of $b$ these 6-d 
metrics preserve the same  fraction  (1/4)
 of type II supersymmetry, so
 that the corresponding
D3-brane solutions should also  have \ci{polg,gub} the same amount of
supersymmetry
 as the solutions of \ci{KT,KS,pt}.
\foot{Preservation of supersymmetry  is obvious
for the pure D3-brane solutions.
For the fractional D3-branes, it was claimed    in 
\ci{cv1,cv2} that the  resolved conifold solution of \ci{pt}
breaks all supersymmetry.}

When $b\not=0$,  the  ``standard"  (no 3-form flux)
 D3-brane on the  conifold solution  has
the  small $\r$ limit   which is
 no longer $AdS_5 \times T^{1,1}$ as  was in  the standard
 $b=0$  case \ci{kw1}. As a result, the
 conformal invariance is broken and
$b$ plays the role of a ``mass scale".
 An interesting feature of
these
solutions is that they  have a curvature
 singularity coinciding  with the horizon at $\r=b$,  like 
in the
``standard" ($b=0$) resolved conifold case 
discussed in \cite{pt} 
(where the singularity and the horizon  were  located at
$\r=0$).

Including  fractional D3-branes changes the situation radically.
 As in
the conifold \cite{KT}  and the resolved conifold  \cite{pt} cases
there is a
naked
singularity of a  repulson type  \ci{repa} 
located  behind the ``zero-charge"  ($F_5=0$) locus.
The gauge theory interpretation of these solutions
should probably go along the lines of the discussion in  \ci{aha}.
It would be very interesting to understand the meaning of the parameter $b$ on the gauge theory side.

Since the simplest (though non-supersymmetric \ci{romans})
 Ricci flat  6-d space  with the same topology
$R \times S^2 \times S^3$   is the cone over $ T^{1,0} = S^2 \times
S^3$,  we
shall, for comparison,    consider also  the corresponding D3-brane
solution.

In section 2 we shall present the explicit form of the
 Ricci-flat metrics with  the $R\times S^2\times
S^3$ topology  referred to above.
 In section 3 we shall construct the
generalizations of the standard  D3-brane solution
 \ci{horst,dul} to the
case when
the transverse 6-d spaces are the
 generalized  ($b\not=0$)  standard, resolved and deformed 
conifolds.

The fractional D3-brane solutions which 
are the $b\not=0$ analogs of  the conifold and
resolved
conifold backgrounds of \ci{KT,pt} 
 will be  found  in section 4.
  In section 5  we shall study, following \ci{maldacena,rey}, 
 the
energy of a  static fundamental string (with both ends at the
``boundary"
$\r=\infty$) in these backgrounds and argue that the corresponding
Wilson  loop
has  confining (area law)
 behavior.
This conclusion seems robust since the ``bent" string  does not reach
the singular
region.


\section{Ricci-flat metrics with topology $R\times S^2\times S^3$}

\label{ricci-flat}

One natural way to construct Ricci flat spaces of topology $R\times
S^2\times S^3$
is to consider cones over Einstein spaces with topology $S^2\times S^3$.
Examples
of the latter are $T^{1,1}$ (supersymmetric)  and $T^{1,0}$
(non-supersymmetric).
 In what follows we shall start with these simplest examples
and consider their natural generalizations.


\subsection{Cone over $S^3\times S^2$}

$T^{1,0}$ is not only $S^3\times S^2$ topologically,
 but geometrically too.
 The resulting 6-d
Ricci-flat metric  is (the radii of the two spheres are adjusted to make
the whole
space
 Einstein one)
\be ds_6^2=d\r^2 + \r^2 \big[{1\over 8}\left(e_{\psi_1}^2  +
e_{\theta_1}^2+e_{\phi_1}^2\right)+{1\over 4}
\left(e_{\theta_2}^2+e_{\phi_2}^2\right)\big] \  \la{conn}, \ee where
the vielbein
is
\be e_{\theta_i}=d\theta_i, \,\,\, \ \  e_{\phi_i}=
\sin\theta_id\phi_i,\ \ \,\,\
e_{\psi_1}=d\psi_1 + \cos\theta_1d\phi_1\ . \la{vei} \ee


\subsection{Conifold  }

The standard conifold metric, i.e. the metric of the cone over $T^{1,1}=
{SU(2)
\times SU(2) \ov U(1) }$ is  \cite{candelas,page}
\begin{equation}
ds_6^2  =d\r^2+ \r^2 \bigg[ \de  e_{\psi}^2 + \si
\left(e_{\te_1}^2+e_{\p_1}^2 +
e_{\te_2}^2+e_{\p_2}^2\right)\bigg] \ , \la{sta}
\end{equation}
   where  $   e_{\te_i}$  and $e_{\p_i}$
are the same as in \rf{vei} and
\begin{equation}
 e_{\psi} = d\psi +  \cos \te_1 d\p_1  +  \cos \te_2 d\p_2  \ . \la{vie}
\end{equation}
To get  a  more general class of Ricci flat
 metrics on the conifold let us
recall some basic relations    from \cite{candelas,minasian,pt,papa2}
(where
 further details and notation may be found).
The conifold can be described as a  quadric in ${\bf C}^4$: $
\sum_{i=1}^4
w_i^2=0$, or  a solution of $ \mbox{det} \W=0$. In terms of the $2
\times 2$ matrix
$\W$ and the  potential $K$ the  K\"ahler metric on the conifold is
($r^2=\mbox{tr}
(\W^{\dagger} \W) =\sum_{i=1}^4 |w_i|^2$)
\begin{equation}
\label{con}
ds^2=K'\mbox{tr}(d\W^{\dagger}d\W)+K''|\mbox{tr}(\W^{\dagger}d\W)|^2 ,\
\ \ \ \ (...)'\equiv  {d\ov d r^2}(...) \ .
\end{equation}
The Ricci tensor for a K\"ahler metric is $R_{p{\bar q}}=-\pd_p\pd_{\bar
q}\ln
\mbox{det}g$, where  for the metric in (\ref{con})
\begin{equation}
\mbox{det}g={1\over |w_4|^2}(K')^2r^2(K'+r^2K'')\ .
\end{equation}
The Ricci-flatness condition  implies
\begin{equation}\label{gam1} [(r^2 K')^3]'=2r^2  \ ,
\end{equation}
 which is integrated to give
\begin{equation}
\label{gam2}   (r^2 K')^3 =r^4+ c\ .
\end{equation}
We shall assume that the constant $c$ is non-negative to avoid a curvature singularity at finite $r$.

 The conifold metric  \rf{con}
 then  is \footnote{The
coordinate $r$ here was denoted  $\rho$ in \cite{papa2}.} \be ds^2 =
{2\over
3}      ( c+ r^4)^{-2/3}  ( r^2 dr^2 +   { 1 \ov 4}
   r^4
e^2_\psi  )
    +  {1\over 4} ( c+ r^4)^{1/3}
(e_{\te_1}^2+e_{\p_1}^2+e_{\te_2}^2+e_{\p_2}^2) \ . \la{cone} \ee For $c> 0$,
writing the metric in terms of $z=\ha r^2$ we get   near $r=0$ \be
(ds^2)_{r\to 0}
=   {2\over 3}      c^{-2/3}  (  dz^2 +   z^2  e^2_\psi  )
    +  {1\over 4} c^{1/3} (
e_{\te_1}^2+e_{\p_1}^2+e_{\te_2}^2+e_{\p_2}^2 ) \ . \la{lii} \ee 
Thus (i) 
 near the apex $(r=0)$  the 2-cycles  stay finite, and (ii)
it is possible to avoid the conical curvature singularity at $r=0$ by
changing the
range of $\psi$  from the original one $[0, 4\pi)$ to  $[0, 2\pi)$.

 The generalized conifold  with $b\not=0$ \rf{cone}
is the 6-d analogue of the Eguchi-Hanson metric  \cite{eh1}
 where to avoid the singularity
one is to change $S^3 \to S^3/{\bf Z}_2 ={\bf P}^3$.  Here
 $\psi$ is the coordinate of the fiber of the $U(1)$
bundle over $S^2\times S^2$. Taking $\psi \in [0, 2\pi)$ one finds
that for
large $r$  the space is now  the cone over
 $T^{1,1}/{\bf Z}_2$;  this is an example of asymptotically locally
Euclidean metric.

To establish the analogy with the Eguchi-Hanson metric  more
explicitly
let us  define  the constant $b\geq 0$ by \be c\equiv ({ 2 \ov 3} b^2)^3
\la{ccc}\
, \ee and introduce the  new radial coordinate $\r$  through the
relation \be
\rho^6=
  [{ 3 \ov 2}  r^2
 K'(r)]^3 =   ({ 3 \ov 2})^3 r^4 +  b^6   \ . \la{ree}\ee 
Since  $0\le r < \infty$  the range of variation of $\r$ is $b\leq
\rho
<\infty$. 

The metric \rf{con},\rf{cone}  then  becomes
\begin{equation}
\label{reso} ds_6^2={ \k^{-1}(\r)}d\r^2+ {1\over 9} \k (\r) \r^2
e_{\psi}^2 +
{1\over 6}
\r^2\left(e_{\te_1}^2+e_{\p_1}^2+e_{\te_2}^2+e_{\p_2}^2\right) ,
\end{equation}
where
\begin{equation}
\k(\r)= 1-{b^6 \over \r^6} \ . \la{kep}
\end{equation}
Note that $ 0\leq  \k  \leq 1 $.
  The analysis  of this metric follows closely
that of
 the Eguchi-Hanson metric in \cite{eh1,eh2}.
It is straightforward to establish that this Ricci-flat
 metric does not have
any scalar curvature singularity, e.g., $R_{ijkl}R^{ijkl}=96 \r^{-16}
(\r^{12}+20b^{12})$ is finite for $ 0< b \le \r < \infty$, 
so that introducing the parameter $b$ smoothens  the 
curvature singularity of the original conifold metric.
In that sense the  $b$-generalized conifold  introduced  above 
may be called ``regularized conifold" (by analogy with resolved and deformed
conifolds which also contain an extra parameters  eliminating the curvature singularity).

 The point
$\rho=b$ is a
removable  bolt singularity,   as can be seen by introducing the
coordinate
$u^2={1\over 9}\r^2 \k(\r) $ and considering the  $\rho\to b$ limit.


\subsection{Resolved conifold  }

Following \cite{candelas,pt,papa2}, one finds the analogous one new
parameter ($b$)
extended metric on  the { resolved conifold} (cf. \rf{reso}) \be ds^2={
\k^{-1}(\r)}d\r^2+ {1\over 9} \k (\r) \r^2 e_{\psi}^2 + {1\over 6}
\r^2\left(e_{\te_1}^2+e_{\p_1}^2\right)+{1\over 6}
 ({\r^2} +
6a^2)\left(e_{\te_2}^2+e_{\p_2}^2\right)\ . \la{resol} \ee Here $a$ is
the
resolution parameter and \be \k(\r)= {1+{9a^2\over \r^2}-{b^6\over
\r^6}\over
1+{6a^2\over \r^2} } \la{kaap} \ . \ee 
 For
$a=0$  this metric reduces to the above conifold metric \rf{reso},\rf{kep}.
 For
 $\r$ much greater than any of the two length scales  $a$ and
$b$  we get the   standard conifold metric  \rf{sta}.

The coordinate  $\r$ and the
original
coordinate $0\le r< \infty $ of
 the conifold
(see \cite{pt,papa2}) are related according to (cf. \rf{ree}) \be
\r^6+9a^2\r^4=({3\over 2})^3 r^4+ b^6. \la{rei} \ee The range of $\r$ is
thus $\r_0
\le \r <\infty$, where $\r_0=\r_0(a,b) \geq 0 $ corresponds to the apex $r=0$, i.e.
is the
solution of $\r_0^6+9a^2\r_0^4-b^6=0$ which is positive  and becomes
zero   when
$b=0$.\foot{Explicitly, $\r_0^2=9a^4\nu^{-1/3}+\nu^{1/3}-3a^2$, where
$\nu=\ha
\big[b^6-54a^6+b^6\sqrt{1-108{a^6\ov b^6}}\big]$ (see
\cite{candelas,pt}).} We
shall
assume   again that $\psi \in  [0, 2\pi)$ to avoid  the conical bolt
singularity at
$ \r=\r_0$.

The  curvature invariants for the metric \rf{resol}
are regular (unless both $a$ and $b$ are zero when we get back to 
the original singular conifold metric \rf{sta}).  In particular, 
$$R_{ijkl} R^{ijkl} = {96 \ov  \r^{12} (\r^2 + 6 a^2)^{6}}
\bigg[ b^{12} ( 5184 a^8  +4320 a^6  \r^2  + 1440 a^4 \r^4
+240 a^2  \r^6  +20  \r^8)  $$
$$ +\   144  b^6  a^4 \r^{10}
+  \r^{12} (6480 a^8   +2160 a^6 \r^{2}
 +360 a^4  \r^{4} + 30 a^2  \r^{6} + \r^{8}) 
 \bigg] \ , 
$$
which is non-singular at $\r\to \r_0$
(for $a=0$ this  expression reduces to the one  conifold one 
given at the end of 
the previous subsection, and for $b=0$ it givbes the  
curvature invariant for the resolved conifold metric of \ci{pt}).

To understand the  short-distance
 $\r \to \r_0$ ($r\to 0$)  behavior  of the $b\not=0$ 
resolved conifold metric 
we introduce as in \rf{lii}
the coordinate $z= \ha r^2 \to 0$  thus getting 
\be (ds^2_6)_{r\to 0} ={3 \ov 8 \r^2_0 (\r^2_0 + 6 a^2) }
\left(dz^2 +z^2e_{\psi}^2\right)
+ \si
\r^2_0\left(e_{\te_1}^2+ e_{\p_1}^2\right)+\si (\r^2_0+6a^2)
\left(e_{\te_2}^2+e_{\p_2}^2\right)\ . \la{smii} \ee 
For fixed $\te_i$ and
$\p_i$   the metric   is thus proportional to   
$dz^2+z^2 d\psi^2$,
so that  to avoid a   conical
 singularity  we  need again to
apply a ${\bf Z}_2$ identification to 
 $\psi \in [0,4\pi)$.\foot{Note that in the $\r\to
\r_0$  limit
 the metric is topologically the same as
that of the  generalized conifold \rf{lii}, i.e. ${\bf R}^2 \times S^2
\times S^2$, but 
geometrically the two metrics are   different (e.g., the 
radii  of the
two
2-spheres   spanned by $(\te_i, \p_i)$ are different).}

The short-distance limit  of the $b=0$ resolved
 conifold metric  (which is to be considered separately 
as $\r_0 =0$ for $b=0$) is \ci{pt} ($ \r^2 \approx \sqrt{3\ov 8a^2}  r^2$)
\be (ds^2_6)_{\r\to 0} ={2 \ov 3}
d\r^2 + \si \r^2 ( e_{\psi}^2
+ e_{\te_1}^2+ e_{\p_1}^2)+   (a^2 + \si \r^2)
(e_{\te_2}^2+e_{\p_2}^2)\ . \la{mii} 
\ee 
This metric  has regular curvature invariants, e.g., 
$(R_{ijkl} R^{ijkl})_{\r\to 0}  \to  { 40 \ov 3 a^4}$, 
illustrating  that the parameter $a\not=0$   indeed  resolves the
curvature singularity of the standard  conifold.


\subsection{Deformed conifold}

The $b$-parameter family of metrics on the deformed conifold is found in
a similar
way \cite{candelas,minasian,papa2,pt}. Using the basis of
\cite{minasian,KS} the
metric can  be written as ($g_5= e_\psi$)
\begin{equation}
\la{ddd} ds^2_6 = \k^{-1}(\r) d\r^2+ {1\over 9} \k (\r) \r^2 e_{\psi}^2
+ {1\over 6}
\r^2\bigg[ \sqrt{r^2 -\ep^2\over r^2+\ep^2}(g_1^2+g_2^2) +\sqrt{r^2
+\ep^2\over
r^2-\ep^2}(g_3^2+g_4^2) \bigg], \la{defo}
\end{equation}
where  $\r$ is related to $r$ according to 
 (cf.  \rf{ree},\rf{rei}) 
\be \label{kkk}
 \r^6 = ({ 3 \ov 2})^3 r^4 \bigg[
\sqrt{1 -{\ep^4\ov r^4}}-{\ep^4\ov r^4} \ln{r^2+\sqrt{r^4-\ep^4}\over
\ep^2}\bigg]   +
b^6 \ , \ee 
and  (cf. \rf{kep},\rf{kaap}) 
\be\la{kuk} \k(\r)=({ 3 \ov 2})^3 {  r^4\ov \r^6} (1 -
{\ep^4\ov r^4})
   = 1 - { b^6\ov \r^6} + O(\ep) \   .
\ee Since  $\ep \le r \le \infty$ we have $b \le \r < \infty$. This
metric reduces
to the generalized conifold metric
 \rf{reso} for $\ep\to 0$. For $r$
greater than any of the two length scales $b$ and $\ep$, it  becomes the
standard
conifold metric \rf{sta}.

It is useful also to present  the deformed conifold 
metric in terms of  the   radial coordinate $\tau$\  used in 
\ci{minasian,KS,pt} 
\be
r^2 =\ep^2 \cosh \tau \ , \ \ \ \ \ \ 
\r^6 =  ({ 3 \ov 2})^3 \ep^4 [\ha  \sinh(2\ta)- \ta]  + b^6 \ , 
 \ \ \ \ \    \  0\le\tau< \infty \ , 
  \ee
 \be 
ds^2_6
=  \ha \KK
\left[
(3\KK^3)^{-1}\ep^4(d\ta^2+g_5^2)
+
{\rm sinh}^2{\tx {\ta\over 2}}\ (g_1^2+g_2^2)
+{\rm cosh}^2{\tx {\ta\over 2}}\ (g_3^2+g_4^2)\right]\ ,  \la{defor}
\ee
where
$$
\KK(\ta)={[c+ \ha \ep^4(\sinh(2\ta)- 2\ta)]^{1/3}\over \sinh\ta}
=  {2\ov 3} \r^2  (\sinh\ta)^{-1}  \  , \ \ \ \ 
\ \  \ c=({2\ov 3})^3 b^6  \ . $$
For large $\tau$ we again  get  the standard  conifold metric, while 
for small values of $\tau$ we get (for $b\not=0$)  
\be  (ds)^2_{\tau \to 0}={3\ep^4\over
8b^4 }\ta^2(d\ta^2+g_5^2)+{b^2\over
12}\ta(g_1^2+g_2^2)+{b^2\over
3}{\ta}^{-1} (g_3^2+g_4^2)\ ,   \la{hieh} 
\ee
or, in terms of $\r \to b$
\be
\label{heeh}
(ds^2_6)_{\r \to b} = \k^{-1}(\r) d\r^2+ {1\over 9} \k (\r) \r^2
e_{\psi}^2 +
{1\over 6} \r^2\big[ ({\r^6 -b^6\ov 18 \ep^4})^{1/3} (g_1^2+g_2^2) +
({\r^6
-b^6\ov 18 \ep^4})^{-1/3}   (g_3^2+g_4^2) \big],
\end{equation}
where for $\k= \k_{\r \to b}= [({3 \ov 2})^5 \ep^4]^{1/3}
{(\r^6-b^6)^{2/3}\over
\r^6}  \to 0 $.  
 Note that the  volume of the 2-cycle $(g_1,g_2)$ shrinks to
zero, while volume of the 3-cycle $(g_3,g_4,g_5)$ stays constant
(=${\ep^2\over \sqrt { 6} }$).
This metric  has a curvature  singularity    at $\tau=0$ for any $b\not=0$.

For comparsion,  the small $\tau$ limit of  the standard 
$b=0$ deformed
conifold metric is   \ci{minasian,KS}
 $$
 (ds)^2_{\tau \to 0}= ({ \ep^4 \ov 12})^{1/3} \bigg[ 
\ha (d\ta^2+g_5^2)  +  g_3^2+g_4^2
  + \fo  \ta^2 (g_1^2+g_2^2)\bigg] \ ,  $$
and is regular at $\tau=0$.
Indeed, computing the curvature invariant 
$R_{ijkl}R^{ijkl}$ for the  general form of the metric \rf{defor}
and then expanding it near $\tau=0$  (i.e. $r=\ep$) 
one finds
a singular expression for $b\not=0$\foot{Here 
$\r$ and $b$ have canonical length $l$ dimensions  while  
$ r\sim l^{3/2}$, $ \ep \sim l^{3/2}$, $\ta \sim l^0$.
Note that the leading term in the expression below is what one finds directly from  the metric \rf{hieh}. } 
$$
(R_{ijkl}R^{ijkl})_{\tau \to 0}={5120\, b^8 \over 9\ep^8}{1\over
\tau^8} \bigg[ 1 - { 8 \ov 15} \ta^2  + { 3 \ep^4 \ov 4 b^6} \ta^3 + O(\tau^4)
\bigg] \ ,  
$$
and a  regular expression for $b=0$:
$$
(R_{ijkl}R^{ijkl})_{\tau \to 0}={192\cdot  (18)^{1/3}  \over 5\,
\ep^{8/3}} \bigg[ 1 
-{4\over 5}\tau^2 +
{202\over 525} \tau^4 + O(\tau^5)\bigg] \ . 
$$


\subsection{Remarks}

 The three
one-parameter families of the  Ricci-flat metrics presented above
\rf{reso},
\rf{resol} and \rf{defo}  are the most general solutions (with
non-trivial dependence on the radial
coordinate   only) for  Ricci-flat metrics on respective conifolds
\ci{pt,papa2}. As was shown in \ci{papa2}, they define supersymmetric
backgrounds
of type II supergravity. This is in contrast to what happens in the case
of the
metric of a cone over $S^3 \times S^2$  \rf{conn} which breaks all
supersymmetries
and thus may  lead to unstable D3-brane solutions.

The analogy with the Eguchi-Hanson metric elucidates  the geometrical
meaning of
the ``mass" parameter $b$. From the perspective of dual gauge theories
associated
with   D3-brane solutions on  these
  generalized conifolds which will be discussed below
this parameter should
 play the role of an IR  ``mass" or ``confinement"  scale.

A  feature of the conifold \rf{reso} and the resolved conifold
\rf{resol} is that
near the apex  $(r=0)$ the respective metrics  effectively ``factorize"
into
 ${\bf R}^2\times S^2 \times S^2$ part.
This will have  consequences  for  the structure of D3-brane solutions
in the IR
region.

We have seen that  while the $b=0$ conifold 
 had curvature singularity at the origin, 
its    $b\not=0$  generalization  has  regular curvature
(and have no conical singularity after changing the period of the 
angle $\psi$).
The  resolved conifold metric depending on two  parameters 
$(a,b)$  is regular for  all of their values except $a=b=0$.
Curiously, this is different from the situation
for the deformed conifold:
   while the standard 
  $b=0$ deformed conifold metric was regular (with the parameter $\eps$
playing the role of the ``cutoff"),  
 its $b\not=0$ generalization has a 
curvature singularity at the origin $r=\eps$. 
This implies, in particular, that one is unlikely 
to find a  direct  $b\not=0$ 
generalization of the regular fractional D3-brane  
on deformed conifold solution of \ci{KS} (which is thus
a very special point in the parameter space of solutions). 
For that reason in what follows  we shall
 mostly concentrate on the resolved conifold case.

Finally, let us note that the conifold solutions  discussed 
above can be derived as special cases of the following ``interpolating"
 ansatz for 6-d metric  (see \ci{papa2})
\be
ds^2_6 =      e^{ 2z + 3 x}  du^2  + 
e^{z+x} [e^{g }  (e_{\theta_1}^2 + e_{\phi_1}^2)     
      +   e^{-g}  ( \tilde \ep_1^2 + \tilde \ep_2^2) ] 
+ e^{-2z -  x} e_\psi^2
\ , 
\la{inte}
\ee
where $\tilde \ep_1 = \ep_1 - a(u) e_{\theta_1}, \ 
\tilde \ep_2 = \ep_2 +  a(u) e_{\phi_1}$, \ 
$ \ep_1=\sin\psi\sin\theta_2 d\phi_2+\cos\psi d\theta_2,$ \
$\ep_2= \cos\psi\sin\theta_2 d\phi_2 - \sin\psi d\theta_2$, 
and $e_{\theta_1}, e_{\phi_1}, e_\psi$ are as in \rf{vei},\rf{vie}.
 Resolved  and deformed conifold metrics are special cases of this ansatz
\ci{papa2} corresponding to $a=0$ and $a^2= 1-e^{2g}$
respectively. 
The unknown functions $x,z,g,a$ 
of the radial coordinate $u$  representing   Ricci-flat 6-d spaces 
are subject to equations following from the following 1-d action (plus the ``zero-energy" constraint)
\be\la{acti}
S_1 = \int du [ x'^2 - z'^2 - g'^2 - e^{-2g} a'^2   - V (x,z,g,a)]
\ , \la{syss} \ee
\be
V=  \ha  e^{-2z} [ e^{2g}  + (a^2-1)^2 e^{-2g}]  -  4 e^{z+2x} \cosh  g   
+ a^2 (e^{- z} - e^{2z + 2x - g})^2  
\ . \ee
For the special cases 
  corresponding to  the resolved and the deformed conifolds
one finds that this system admits a superpotential $W$ \ci{pt}, 
i.e. $ V= - G^{ij} {\del W\ov \del q^i } {\del W\ov \del q^j}, $ where 
$ q^i=(x,z,g,a)$,\  $ G_{ij} = (1,-1,-1,-e^{-2g})$, 
so that one gets a first-order system
$q'^i= G^{ij} {\del W\ov \del q^j}$. For example, 
in the resolved conifold case ($a=0$)\  
 $W= e^{-z}\cosh g + e^{2z+2x}$.
It would be very interesting to find new solutions of the sytem \rf{syss} 
representing 6-d Ricci-flat spaces that  ``interpolate" between the 
generalized  resolved and deformed conifold  metrics
discussed above. dd


\section{Pure D3-brane solutions}

\subsection{General remarks}
As is well known, given a Ricci flat 6-d space with
 metric $g_{mn}$,
one can construct the following generalization of the standard
\cite{horst,dul}
 3-brane solution (see, e.g.,
\cite{kehagias,minasian,papa1}) \be
ds^2_{10}=h^{-1/2}(y)dx^{\m}dx^{\m}+h^{1/2}(y)g_{mn}(y)dy^{m}dy^{n}\ ,
\la{meet}\ee
\be F_5=(1+*)d h^{-1}\wedge dx^0\wedge dx^1\wedge dx^2\wedge dx^3 \ , \
\ \ \ \ \
\Phi=\const\ , \la{typ} \ee where  $h$ is a  harmonic function  on the
transverse
6-d space:
\begin{equation}
\label{d3eqn} {1\over \sqrt{g}}\pd_m\left(\sqrt{g}g^{mn}\pd_n h\right)=
0\ .\la{har}
\end{equation}
In the case of a Ricci-flat cone with the metric
\begin{equation}
g_{mn}(y)dy^{m}dy^{n}=d\r^2+\r^2 \g_{ij}(z) dz^idz^j \ ,
\end{equation}
one can choose $h$ to depend only on $\r$, thus getting the
single-center solution
\begin{equation}
h =h_0 +{L^4\over\r^4}\ . \la{coned}
\end{equation}
Here and below we assume that $h_0 \geq 0$. Then in the near-core region
the space
becomes  $AdS_5\times X^5$, where $X^5$ is the Einstein space which is
the  base of
the cone with the  metric $\g_{ij}$.
Particular examples
 are provided by  the standard cone metrics
\rf{conn} and \rf{sta} where
 $X^5=T^{1,0}=S^2\times S^3$  and $X^5=T^{1,1}$.\footnote{Compactifications of IIB
supergrativity
 on such $X^5$
were discussed in \cite{romans} where it was pointed out that in  the
family of
$T^{p,q}$ spaces only $T^{1,1}$ preserves
 supersymmetry.}

In the general case when the transverse 6-space is not a cone, one  will
not find
the near-core geometry having $AdS_5$ factor and thus the dual gauge
theory will
have broken conformal invariance.

\subsection{Conifold case}
 Let us determine
the harmonic function $h$ that solves  \rf{har} for the generalized
conifold metric
 (\ref{reso})
  assuming
  $h=h(\r)$. Using that
$ \sqrt{g_6}= {1 \ov 108} \r^5\sin\te_1\sin\te_2, $ $
g^{\r\r}=1-{b^6\over \r^6} $
we get
\begin{equation}\la{hh}
h=h_0 -{2L^4\over b^4}\left[\si \ln{({\rr^2} -1 )^3\over { \rr^6} - 1
}+{
{ 1\ov \sqrt{3}}}({\pi\over 2}-\arctan{2{\rr^2} +1\over
\sqrt{3}})\right],
\end{equation}
$$ \rr \equiv {\r \ov b} \ . $$
For ${\r\over b}\gg 1$ we
 recover the standard metric of
D3-branes on the conifold \rf{coned}, i.e. \be h(\rho \to \infty)
=h_0+{L^4\over
\r^4}\ .\ee
 For small values of  $\r$  we get $h\to \infty$
 \be h(\r \to b)
=-{2L^4\over 3
b^4}\ln\left({\r\over b}-1\right) +h_1 \ , \ \ \ \ \ \ \ \ \ h_1=h_0 + {
L^4\over 9
b^4}(\sqrt{3}\pi+3\ln{4\over 3}) \ . \la{ggg} \ee Recalling from
\rf{meet} that the 00-component of the metric is 
$h^{-1/2}$ we see that $\rho=b$ is a horizon. It is also a
curvature
singularity. Thus, starting with   ${\bf R}^{1,3} \times M^6$ where
$M^6$ is the
generalized  $b\not=0$  conifold,
 the introduction of D3-branes creates
a back reaction that transform the previously nonsingular point $\r=b$
into a
curvature singularity coinciding with
 horizon.
 In contrast, for $b=0$ \cite{kw1}  the near-core geometry was
regular -- $AdS_5\times T^{1,1}$.

This  is similar to  what was found for the D3-brane on  $b=0$
 resolved ($a\not=0$) conifold  in  \cite{pt}.
Below we will  see that this behavior extends  also to the case of the
resolved
conifold with  $b\not=0$.

One of the generic features of D3-branes on the conifold and the
resolved conifold
is the logarithmic form 
  of  $h(\r)$ at small distances. The origin of
this
 lies in the fact (noted in section 2.5)
that for small $\r$ the transverse geometry  becomes effectively
2-dimensional as
far as the dependence on the radial coordinate is concerned, so that one
finds the the harmonic function $h$ has 
 the  ``7-brane-like"  $\log (\r-\r_0)$
  structure.


\subsection{Resolved conifold case}

Starting with the
 { resolved conifold}  metric \rf{resol} and solving \rf{har}
one finds the following expression for  $h(\r)$
\begin{equation}\la{hhh}
h=h_0-{2L^4\over \alpha b^4}\left[ \si  \ln{ ({\rr^2} - {\rr_0^2} )^3
\over {\rr^6}
+ 3 q^2{\rr^4} - 1 } + {\beta+q^2\over \lambda} ({\pi\over
2}-\arctan{2\rr^2+\sigma
\over \lambda})\right], \ee
 where as in \rf{hh} 
($\r_0$ is the same as in section 2.3)
 $$ \rr = {\r\ov b}\ ,
\ \ \ \ \ \ \  \ \rr_0 =  { \r_0\ov b}  \ , $$  and for simplicity of
presentation
we have introduced the following constants depending on the ratio $a/b$
\begin{eqnarray}
q=\sqrt{3}{a\over b}, \qquad
 \beta = q^4\mu^{-1}+ \mu- q^2, \qquad
\mu\equiv [{ 1 \ov 2} (1-2q^6+\sqrt{1-4q^6})]^{1/3},  \nonumber \\
\alpha= { 1 \ov 3} (2\beta^2+3q^2\beta +\beta^{-1}), \qquad \sigma =
\beta +3q^2,
\qquad \lambda = \sqrt{4\beta^{-1}-\sigma^2}. \la{parr}
\end{eqnarray}
We have assumed that  ${1-4q^6} >  0$ (the opposite  case is
discussed
below).

The behavior for  $ \r\to \r_0$ is the same as in \rf{ggg} \be
h(\r\to\r_0)=-{2L^4\over 3\alpha b^4}\ln({\rr} - {\rr_0} ) +O(1) \ , \ee
i.e.,  as
in the previous case,  at $\r=\r_0$ there is a horizon coinciding  with 
curvature
singularity.

Compared to \rf{hh} here the   function $h$
 contains another (resolution)  scale $a$ represented by the
parameter $q \in [0,\infty)$.
  Depending on the  value of $q$, i.e. on the 
 ratio  of $a$ and $b$,  
one may distinguish
 three  regions: $b$-dominated, intermediate, and  $a$-dominated.
 The 
 metric defined by 
 \rf{hhh},\rf{parr} 
is valid  in the $b$-dominated region
where  $4q^6<1$, i.e., ${b}> 3^{1/2}2^{1/3} a$. For $q=0$
($a=0$)  we
get  back to the conifold case \rf{hh} which should be viewed as the
limiting case
of the $b$-dominated resolved conifold
solution  \rf{hhh}. In the intermediate
 region, i.e.
$4q^6=1$ (this corresponds to $\lambda =0$ in \rf{hhh},\rf{parr})
  we get
\be h=h_0-{2L^4\over 81a^4}\left({9a^2\over
\r^2+6a^2}+\ln{\r^2-3a^2\over
\r^2+6a^2}\right), \ee where $\sqrt{3}a\le \r <\infty$. 
This metric also
has a
curvature singularity and  the horizon at
$\r=\r_0=\sqrt{3}a$. In the
$a$-dominated region $4q^6>1$, i.e.
${a}> 3^{-1/2}2^{-1/3} b$, 
 we find \be h=h_0-{2L^4\over 9
a^4\alpha}\left[
\si  \ln{ ({\rr^2} - {\rr_0^2} )^3 \over {\rr^6} + {\rr^4} - q^{-6} }
-{\beta+1\over \lambda} \ln{2\rr^2+\sigma -\lambda \over 2\rr^2+\sigma
+\lambda}\right], \la{aaa} \ee where\footnote{ Note that since
$|\mu|=1$, the
combination  $\mu+\mu^{-1}$ is always real so that  $\beta,\, \sigma,\,
\lambda$
and $\alpha$ are always real and $\r_0$ is positive real root of the
cubic
equation.}
\begin{eqnarray}
\rr={\r\ov \sqrt{3}a} \ ,  \qquad
 \beta = \mu+\mu^{-1} -1, \qquad
\mu\equiv e^{i\pi/3}[1-(2q^6)^{-1} -{i q^{-3}}\sqrt{1-
(4q^6)^{-1}  }]^{1/3},  \nonumber \\
\alpha= { 1 \ov 3} [2\beta^2+3\beta +(q^6 \beta)^{-1}], \qquad \sigma =
\beta +3,
\qquad \lambda = \sqrt{\sigma^2-4(q^6 \beta)^{-1}}.
\end{eqnarray}
This solution has the same generic logarithmic behavior near $\r_0$,
indicating the
existence of a horizon and a singularity.

In $a$-dominated expression 
 \rf{aaa} we  are able to take $q \to \infty$
($b\to 0$)  limit
to
get back to the ``standard"
 resolved conifold case. The function $h$
in the metric of D3-branes on $b=0$ resolved conifold   found  in \cite{pt}  is
\be
h_{b=0}   =h_0 + { 2 L^4 \ov 81 a^4} \bigg[ { 9 a^2\ov  \r^2} - \ln (
1 +
 { 9 a^2 \ov \r^2})\bigg] \ .
 \la{resold}
\ee Again, here 
 $\r=0$ is both the  horizon and the  singularity.


\subsection{Deformed conifold case}

In  the deformed conifold case \rf{defo},\rf{kkk}
 the harmonic function $h$ in \rf{meet} is  found to be
\begin{equation}
h=h_0      - { 2^5\ov 3^3}  L^4   \int {\r d \r\over r^4-\ep^4}\  ,
\end{equation}
where the   explicit form of $r(\r)$ in
\rf{kkk} is
transcendental.  For $r\gg \ep$ we have $r^4=(2\r^2/3)^3$ and therefore
recover the
D3-brane on the conifold metric with $h= h_0 + { L^4\ov \r^4}$. For
$r\to \ep$
($\r\to b$) we get \be ds_{10}^2=h^{-1/2}dx^\m dx^\m + h^{1/2}
(ds_6^2)_{\r\to b} \
, \ee where $(ds_6^2)_{\r\to b}$ is  given by  \rf{heeh} and
\be h=h_1 - h_2 \r^2
+ O(\r^4) \ , \ \ \ \ \ \ \ \ \ h_2 = {2^{8/3}L^4\over 3^{5/3}b^2
\ep^{4/3}}\
{}_2F_1({1\over 3}, {2\over 3}; {4\over 3};{1})\ , \la{deff} \ee 
 and
 $h_1=h_0+ O({L^4\ov b^4})$.
\footnote{Here ${}_2F_1({1\over 3}, {2\over 3}; {4\over
3};{1})={\Gamma({1\over 3})
\Gamma({4\over 3})\over \Gamma({2\over 3})}\approx 1.77$.}
  This $b\not=0$ case  is
different from the  $b=0$ deformed conifold  case
 \cite{KS,pt} where $\r=0$ was  a horizon.  Here for $\r \to b$
the space factorizes into ${\bf R}^{1,3} \times  M^6$  where $M^6$ has
 $\r=b$ as its   curvature singularity.


\section{Fractional D3-brane solutions }

Let us now construct
 the  $b\not=0$ generalization of the fractional
D3-brane on resolved conifold  solution of \ci{KT,pt}, i.e. the
extension of the D3
brane solution of the previous section to the case  of additional
(self-dual)
3-form flux. The resolved conifold solution  includes the conifold one
as a
special ($a=0$) case. The first-order system  corresponding 
to this  background was already obtained  in \ci{pt}.
 It is straightforward also to construct a similar
$b\not=0$
generalization of the 
 solution \ci{KS} in the deformed conifold case (see \ci{pt}), but we
shall not discuss the details of  this here.

For comparison, we shall start with a similar 
case of
3-branes on the cone over $S^2\times S^3$. This solution was
previously
discussed in \ci{ktt} (see also \ci{herz}).


\subsection{$S^2\times S^3$ cone case}

The  3-brane ansatz for the
 metric  with transverse part given by \rf{conn} is
\be ds^2=h^{-1/2}dx^\m dx^\m + h^{1/2}\big(d\r^2+\ei \r^2 d\Omega_3^2
+\fo \r^2
d\Omega_2^2\big) \ ,  \ee and the natural
 ansatz for the form fields is
similar to
the one in the  conifold  \ci{kn,KT} case 
\begin{eqnarray}
 B_2&=&f(\r)e_{\theta_2}\wedge e_{\phi_2}\ \ \  \to\ \ \ \
H_{3}=f'(\r)d\r \wedge
e_{\theta_2}\wedge e_{\phi_2}, \nonumber \\
F_3&=&P e_{\psi}\wedge e_{\theta_1}\wedge e _{\phi_1}, \nonumber \\
F_5&=&{\cal F}+ *{\cal F}, \, \, \, \ \ \ \ \
 {\cal F}=\K(\r) e_{\psi}\wedge
e_{\theta_1}\wedge e_{\phi_1}\wedge e_{\theta_2}\wedge e_{\phi_2}\ .
\end{eqnarray}
The 10-d duals of these fields are 
\be 
{*}{\cal F}={2^{13/2}  \K
\over\r^5  h^2
}d\r\wedge dx^0\wedge dx^1 \wedge dx^2 \wedge dx^3\ , \ee \be
{*}F_3={2^{5/2}P\over
\r h }d\r\wedge dx^0\wedge dx^1 \wedge dx^2 \wedge dx^3\wedge e_{\te_2}
\wedge
e_{\p_2} \ , \ee \be {*} H_{3}=-{\r f'\over 2^{5/2}  h }dx^0\wedge dx^1
\wedge dx^2
\wedge dx^3\wedge e_{\psi} \wedge e_{\te_1}\wedge e_{\p_1} \ . \ee We
shall assume
that the dilaton $\Phi$ is constant.
 Then the  $F_3$ equation
of motion $ d(e^\Phi *F_3)=F_5\wedge H_{3} $ is satisfied automatically,
and from
the $H_{3}$ equation $ d(e^{-\Phi} * H_{3})=-F_5\wedge F_3 $ one
obtains  the
following  equation ($ e^\Phi= g_s$) \be \bigg({f'_1\r\over h}\bigg)'=
{2^9g_s P
\K\over h^2\r^5}\ . \ee The constant dilaton
 condition implies $H_{3}^2=e^{2\Phi} F_3^2$, i.e.
 using \rf{reso} we get\foot{As in
 \ci{KT}, the axion
 equation is satisfied automatically since $H_3\cdot F_3 =0$.}
$ \r f'={2^{5/2}g_sP}.$ The Bianchi identity for the 5-form \ $
d*F_5=dF_5=H_{3}\wedge F_3$ gives $ \K'= P f', $ i.e.
 \  $   \K   = Q + P f.$
The two linearly independent Einstein equations are a consequence of
this system of
first order differential equations. The solution is thus very similar to
the
original conifold one \cite{KT}
\begin{eqnarray}
f&=&2^{5/2}g_sP\ln {\r\over \r_0}\ , \qquad \ \ \
\K=Q + 2^{5/2} g_sP^2\ln {\r\over \r_0}\ , \nonumber \\
h&=&h_0+{2^{9/2}\over \r^4}\bigg[Q+2^{5/2}g_sP^2(\ln{\r\over \r_0}+ \fo
)\bigg]\ .
\end{eqnarray}
Note that as in \ci{KT},
 the complex 3-form  $G_3=g_sF_3+iH_3$ is
selfdual in 6-d sense. Like the conifold 
 solution \cite{KT}, this solution
has a naked
singularity located at $\r=\r_h$, \ \ $ Q+2^{5/2}g_sP^2(\ln{\r_h\over
\r_0}+
\fo)=0$, i.e.  very close to the origin if   the number
 of
fractional D3-branes is small,  ${Q\gg g_sP^2}$. In this case the
singularity is
behind the  ``zero charge" ($\K=0$)  locus.



\subsection{Resolved conifold case}

The ansatz for the  metric will  be the same as in \rf{meet},
 \be
 ds_{10}^2=h^{-1/2}(\r) dx^\m dx^\m +h^{1/2}(\r) ds_6^2\ , \la{mee}
\end{equation}
where $ds_6^2$  will be  the metric of the generalized $b\not=0$
resolved conifold
(\ref{resol}). Our ansatz for the NS-NS 2-form  will be 
 as in
\ci{pt},
i.e. a natural  generalization of the ansatz in \ci{KT}  motivated by an
asymmetry
between the two $S^2$ parts of the resolved conifold metric
$$
B_2 = f_1(\r)e_{\te_1}\wedge e_{\p_1}+f_2(\r)e_{\te_2}\wedge e_{\p_2} \
,
$$
\be \label{fields} H_{3}=dB_2= d\r\wedge[f'_1(\r)e_{\te_1}\wedge
e_{\p_1}+f'_2(\r)e_{\te_2}\wedge e_{\p_2}]\ . \ee The   conifold case
($a=0$)
corresponds  \ci{kn,KT} to  $f_1=-f_2 $.  The forms $F_3$ and $F_5$
 will also  have the same structure  as in \ci{KT,pt}
(we follow the notation of \ci{pt}) \be F_3 = P e_{\psi}\wedge
(e_{\te_2}\wedge
e_{\p_2}- e_{\te_1}\wedge e_{\p_1} )\ , \ee \be \la{fre} F_5= {\cal F}+*
{\cal F}\
, \quad  \ \ \ \ {\cal F} = \K(\r)e_{\psi}\wedge e_{\te_1} \wedge
e_{\p_1} \wedge
e_{\te_2}\wedge e_{\p_2}\ . \ee The rest of the discussion is
essentially the same
as in \ci{pt} with $\k$ in the 6-d metric now being dependent 
also on $b$
according to
\rf{kaap}. 

Assuming  that the dilaton $\Phi$ is constant, the  $F_3$
equation of
motion $ d(e^\Phi *F_3)=F_5\wedge H_{3} $ is satisfied automatically,
and from the
$H_{3}$ equation $ d(e^{-\Phi} * H_{3})=-F_5\wedge F_3 $ one obtains the
following
three   equations ($ e^\Phi= g_s$) \be \bigg({f'_1\r\k\G\over h}\bigg)'=
{324g_s P
\K\over h^2\r^5\k\G}\ , \ \ \ \ \ \ \ \ \ \bigg({f'_2\r\k\over
h\G}\bigg)'=-{324g_s
P \K\over
 h^2\r^5\k\G}\ , \ee
\be f'_1+ \G^{-2}f'_2=0 \ , \ \ \ \ \ \ \ \G \equiv {\r^2   + 6a^2\ov
\r^2 } \ ,
\la{las} \ee where $\G$
 is  the ratio of the squares of the  radii
of the two spheres in the resolved conifold  metric \rf{resol}. The
constant dilaton
 condition implies $H_{3}^2=e^{2\Phi} F_3^2$, i.e.
\begin{equation}
f'^2_1+ \G^{-2}f'^2_2={9 g_s^2P^2\over k^2\r^2} (1+ \G^{-2} )\ .
\end{equation}
Combined with \rf{las} that gives \be f'_1= {3g_sP \over \r\k\G}\ ,
\qquad \quad
f'_2=-{3g_sP\G\over \r\k}\ . \la{fff} \ee The Bianchi identity for the
5-form \ $
d*F_5=dF_5=H_{3}\wedge F_3$  implies
\begin{equation}
\K'= P (f'_1-f'_2)\  , \ \ \  \  {\rm i.e. } \ \
 \ \ \   \K   = Q + P (f_1-f_2)  \ . \la{kek}
\end{equation}
As in \ci{KT,pt},
 to determine the metric function  $h(\r)$  it is sufficient
 to consider the
trace of the Einstein equations,\footnote{More precisely,
 there are two
linearly independent Einstein equations: one is the square of eqn.
(\ref{las}) and
another, written above, can be expressed in terms of the first
derivative of
(\ref{las}) using  (\ref{kek}).} \, $R = - \ha \Delta h = { 1 \ov 24} (
e^{-\Phi}
H_3^2 + e^{\Phi} F_3^2) $, i.e.
\begin{equation}
\label{source}
 h^{-3/2}{1\over \sqrt{g}}\pd_{\r}\left(\sqrt{g}g^{\r\r}\pd_{\r} h
\right) = - {\tx { 1 \ov 12}}  (g_s^{-1} H_{3}^2+g_s F_3^2) = -\si
g_sF_3^2\ ,
\end{equation}
or
 \begin{equation}\la{sou}
\left(\r^5\k\G h' \right)'= - 324 g_sP^2 {( 1+\G^2) \over \r\k\G}\  .
\end{equation}
Integrating this  we get \be h'  =-   {108 \K\over \r^5\k\G} \ . \la{fe}
\la{pio}
\ee 
Plugging in the function  $\k$ in \rf{kaap} the  system  \rf{fff},\rf{fe} of first-order differential equations
can be directly integrated.

Here we shall present  the explicit  form of the solution  only in 
 the  $a=0$
limit, i.e. the $b$-generalized conifold case \rf{kep}.
For $a=0$ one finds 
  $\G=1$,
implying  $f_2=-f_1$, 
 and with $\k= 1 - {b^6\ov \r^6}$  we
 obtain
\be f_1=-f_2= \ha g_sP\ln({\rr^6}-1) +f_0\ , \qquad
\K=Q+g_sP^2\ln({\rr^6}-1), \ee
$$
h=h_0-{54Q\over b^4}\left[ \si\ln{(\rr^2- 1)^3\over \rr^6-1} +{ {1\over
\sqrt{3}}}({\pi\over 2}-\arctan{2\rr^2+1\over \sqrt{3}})\right]
$$
\be +{27g_sP^2\over  b^4 (\rr^6- 1)^{2/3}} \left[\ {  {3\ov 2}}\
{}_3F_2({2\ov 3},
  {2\ov 3},{2\ov 3};{5\ov 3}, {5\ov 3};-{1\over \rr^6-1})
 + \ {}_2F_1({2\ov 3}, {2\ov
3};{5\ov 3}; -{1\over \rr^6-1})\ \ln({\rr^6}-1)\right],\la{hah} \ee
where $\rr= { \r
\ov b}$ (cf. \rf{hh}) and
 ${}_pF_q$ is the  hypergeometric function.

The large $\r$ behavior of $h$ is \be h(\r\to\infty)=h_0+{27\over
\r^4}\bigg[Q+6g_sP^2(\ln{\r\ov b}+\fo)\bigg]\ , \ee 
i.e. this solution 
has the same UV asymptotic  as the $b=0$ conifold  one 
of \ci{KT}. In
the  short-distance  limit
$\r\to
b$ limit  we have, to  the leading order,
 \be
 h(\rr\to 1)=h_0-{18Q\ov b^4}\ln(\rr^2-1) -{9g_sP^2\over
b^4}\ln^2(\rr^2-1)\ .
 \ee
At $ \rr_h\approx 1+e^{-2Q/(g_sP^2)}$ the solution has a naked
singularity of  a
repulson type. The ``zero charge" locus  $(\K=0)$ is located at $
\rr_{\K} =
1+e^{-Q/(g_sP^2)}$, i.e.
 $\rr_{\K}>\rr_h$.

\bigskip\bigskip

We can thus conclude, based on the analysis in  \cite{KT} $(b=a=0)$, in
\cite{pt}
$(b=0,\, a\ne 0)$ and here   $(a=0, \, b\ne 0)$, that 
generically
fractional
3-branes on the  conifold and resolved conifold 
 have a repulson-type naked
singularity
which  is located behind the ``zero-charge"  locus.


\section{Wilson Loop behavior }

\label{wilson}

Let us now investigate, following \cite{maldacena,rey}, the behavior of
the
 Wilson loop
corresponding to ``quark-antiquark" potential in the dual gauge theory. It
is given by
the exponential of the classical fundamental string action in  these
D3-brane
backgrounds evaluated for a static  configuration of open  string ending
on the
probe D3-brane placed at the ``boundary" $\r= \infty$.

We will show that one gets an area law (confining) behavior  for  the ``pure" D3-brane  backgrounds of section 3, 
assuming   at least
one of the scales of  the transverse space is kept non-zero.\ This  is
different
from  what is found  in
 the standard conifold case \ci{kw1}
 where  the near-core geometry
has $AdS_5$ factor and thus the  potential is Coulombic as in
\cite{maldacena,rey} (in the single-center case  as well as in the
multicenter
case  \cite{wilsoncon}).

For simplicity, we shall consider only  the  D3-brane  background
 with  the resolved conifold as the transverse space.
The corresponding metric  \rf{resol},\rf{hhh} depends on the two scale
parameters
$b$ and $a$. Expressed in terms of $\rr= \r/b$ it depends only on their
ratio
$q=\sqrt 3 a/b$. It is sufficient to analyze the  Wilson loop in the two
limiting
cases $q=0$ and $q=\infty$: (i) $a=0, \ b\not=0$, i.e. D3-brane on
generalized
conifold \rf{reso},\rf{hh}, \ and (ii) $a\not=0, \ b=0$, i.e. D3-brane
on
``standard" resolved conifold \rf{resold}. In both special  cases the
scale of the
transverse space ($b$ or $a$) determines the confinement scale. The
behavior of the
Wilson loop for   general values of $q$ will be similar,  given that the
behavior of
$h$ is generic.  Let us  emphasize  
   that  in contrast to other supergravity solutions
dual to
confining $\cal N$=1 gauge theories \cite{pols,KS,MN}, this confinement
behavior is
found for  the pure D3-brane background  which does not have any
non-trivial
 3-form fluxes.

\subsection{General  set-up}

All examples we have  discussed above have metrics of the type \be
ds^2=h^{-1/2}(\r) (-dx_0^2+ dx_kdx_k) + h^{1/2}(\r) [ \k^{-1}(\r) d\r^2+
ds_5^2]\ ,
\ee where $ds_5^2$ is the metric of the corresponding 5-d compact space.
The
Nambu-Goto string action which determines the expression for the Wilson
loop
depends on this 10-d  metric $G_{MN}$ as $\int d\ta d\s \sqrt{-\det
(G_{MN}\del_a
X^M\del_b X^N)}$.
 In the  static
gauge ($x_0= \tau, \ x_1\equiv x= \s$) and assuming that the string is
stretched
only in the radial direction, i.e. only
 $\r$ coordinate  depends
on $\s$,  we get\foot{$T$ is the time interval and  the string tension is
set  equal to
1.} \be S=T\int dx \sqrt{G_{00}G_{xx}+G_{00}G_{\r\r}(\pd_x \r)^2} = T\int
dx
\sqrt{h^{-1}+\k^{-1}(\pd_x \r)^2}\ .  \ee Since the Lagrangian of this
``mechanical
system" does not depend explicitly  on ``time"   $x$, we have a
conserved quantity $
{h^{-1}\over \sqrt{h^{-1}+\k^{-1}(\pd_x \r)^2}}, $ i.e. the first
integral is ($c_0
=\const$) \be dx= {d\r\over \sqrt{\k h^{-1}\left({h^{-1}\over
c_0^2}-1\right)}} \ .
\la{xxx} \ee
 The energy
of a  static string configuration is thus \be E={S\over T}= \int dx
\sqrt{h^{-1}+\k^{-1}(\pd_x \r)^2}= \int {d\r\over \sqrt{\k
\left({1}-c^2_0
h\right)}}\ . \ee Following \cite{maldacena,rey}, 
 the question  about
confinement
is  then
 reduced to finding the dependence of the
energy $E$ on the  distance $\el$ between the string end-points (between
``quark"
and the ``antiquark").

\subsection{Conifold case}

For the  generalized  conifold  metric with the scale $b$
\rf{reso},\rf{kep} the
function  $h$ of D3-brane solution is given by \rf{hh}.
 Introducing the new coordinate \be y=\rr^2=
{\r^2\over b^2} \ , \ee and removing the asymptotically flat region
(i.e. dropping
$h_0$)
 we obtain the following relation  for the quark-antiquark
separation \be {\el \over 2}= {L^2\over
\sqrt{2}b}\int\limits_{y_*}^{\infty} dy {  y
\over \sqrt{y^3-1}}{f(y)\over \sqrt{f(y_*)-f(y)}} \ , \la{uuu} \ee where $y_*$ is
the turning point and \be f(y)={1\over \sqrt{3}}\left(\arctan {2y+1\over
\sqrt{3}}
-{\pi\over 2} \right)-{1\over 6}\ln {(y-1)^3\over y^3-1}\ , \ \ \ \ \ \
f(y_*)={b^4\over 2L^4 c_0^2}\ . \ee Note that for any finite value of
$f(y_*)$ one
has $y_*>1$,  meaning that the minimal surface does not reach the $\r=b$
which is the 
horizon and the curvature singularity. The energy of the string
configuration is \be
E={b^3\over 2^{3/2}L^2 c_0}\int\limits_{y_*}^{\infty}{ydy\over
\sqrt{y^3-1}}{1\over
\sqrt{f(y_*)-f(y)}}\ . \ee Evaluating the integrals as in
\cite{petrini}, i.e.
assuming that the main contribution comes from the region near $y_*$,
  we find the  ``area law", i.e. the
linear confinement behavior \be E\approx {c_0\over 2} \el \ . \la{are}
\ee


\subsection{Resolved conifold case}

Let us first  consider  the ``standard" $b=0$ version of the D3-brane
solution on
resolved conifold \ci{pt}, i.e. \rf{resold}. Introducing  the  new
coordinate \be
y={\r^2 \over 9a^2 } \ , \ee
  and setting $h_0=0$ we get
 \be
h={2L^4\over 81 a^4} f(y) \ , \ \ \ \ \ \ \ \ \ 
f(y) \equiv y^{-1} -\ln(1+y^{-1}) \ ,
 \ \ \ \
 \qquad \k={y + 1 \over y+{2\over 3} }\ .  \ee
 Then the analog of \rf{uuu} is
\be
 {\el \over 2}=
{L^2\over 3\sqrt{2}a }\int\limits_{y_*}^{\infty } {dy\over y^{1/2}}
\sqrt{{y+{2\over 3}\over y+1}}{f(y)\over \sqrt{f(y_*)-f(y)}}\ ,
\la{deno} \ee where
$ f(y_*)={81
a^4\over 2L^4
c_0^2}.$
  We have used that from the form of
the denominator  in the      analogue of eq. \rf{xxx}
      (cf. \rf{deno})
it follows that there is a turning point  for $y$, i.e.  $y$ changes
from $\infty$
($\r=\infty$) to $y_*$.
 Note that  $f(y)$ is a positive
function and it increases monotonically from zero at $y=\infty$.
 Therefore,  for any positive constant
$d_0$ there is  $y=y_*$ that solves $d_0^2=y^{-1} -\ln(1+y^{-1} ) $.

Similar behavior is found when we switch on the $b$-parameter, i.e.
start with $\k$
and $h$ given in \rf{kaap} and \rf{hhh}. Thus the  minimal surface does
not reach
the curvature  singularity
 located at $\r_0$.\foot{For $b=0$ one  can estimate the value of $y_*$
as follows. For small $d_0$ we expand $f(y)$ near zero and find
$y_*=(\sqrt{2}d_0)^{-1} $, i.e. $ \r_*\approx L\sqrt{c_0}\ .  $
 For large $d_0$, we expand $f(y) $ for
large $y$ to find $y_*=d_0^{-2}$, i.e.
 $ \r_*\approx
{\sqrt{2}L^2c_0\over 3a}. $} 
The expression for the energy  is (for $b=0$) \be E= {27
a^3\over
2^{3/2}L^2c_0}\int\limits_{y_*}^\infty {dy\over y^{1/2}}
\sqrt{{y+{2\over 3}\over
y+1 }} {1\over \sqrt{f(y_*)-f(y)}}\ . \ee Assuming that the main
contribution comes
from the region near $y_*$  and expanding $f(y)\approx
f(y_*)+f'(y_*)(y-y_*)$ we
again get  the area law behavior, i.e. the relation \rf{are}.
 Analogous  result is found
when one switches on the dependence of the
 background metric  on
the parameter $b$.

\section*{Acknowledgments}
We are  grateful to  I. Klebanov and G. Papadopoulos for useful
discussions and
comments. LAPZ would like to acknowledge the Office of the Provost at
the
University of Michigan and the High Energy Physics Division of the
Department of
Energy for support. The work of AAT  was  supported in part by the DOE
grant
DE-FG02-91ER40690,
INTAS project 991590 and PPARC SPG grant  PPA/G/S/1998/00613.


\end{document}